\documentclass[twocolumn,preprintnumbers,amsmath,amssymb]{revtex4}
\usepackage{graphicx}
\usepackage{dcolumn}
\usepackage{bm}
\newcommand{\comment}[1]{}
\newcommand\etal{\mbox{\textit{et al.}}}

\begin{document}
\setlength{\unitlength}{0.7\textwidth}
\preprint{}

\title{A single-time two-point closure based on fluid particle displacements}
\author{W.J.T. Bos (correspondent)}
\author{J.-P. Bertoglio}

\affiliation{%
Laboratoire de M\'ecanique des Fluides et d'Acoustique, UMR CNRS 5509 - Ecole Centrale de Lyon\\
69134 Ecully, France \\
}

\begin{abstract}
A new single-time two-point closure is proposed, in which the equation for the two-point correlation between the displacement of a fluid particle and the velocity allows to estimate a Lagrangian timescale. This timescale is used to specify the nonlinear damping of triple correlations in the closure. A closed set of equations is obtained without ad-hoc constants. Taking advantage of the analogy  between particle displacements and scalar fluctuations in isotropic turbulence subjected to a mean scalar gradient, the model is numerically integrated. Results for the energy spectrum are in agreement with classical scaling predictions. An estimate for the Kolmogorov constant is obtained.
\end{abstract}

\maketitle

Two-point single-time closures are efficient and useful tools for studying homogeneous turbulence. Among existing single-time theories, the Eddy Damped Quasi-Normal Markovian theory (EDQNM), originally proposed by  Orszag\cite{Orszag} (see also Leith \cite{Leith2}) for isotropic turbulence, has been used to investigate a broad range of fundamental problems in turbulence. Examples are: scalar decay in isotropic turbulence\cite{VignonPOF,Herring}, rotating turbulence\cite{CambonJacquin}, stratified turbulence\cite{GodeferdPOF}, magneto-hydrodynamics \cite{PouquetMHD}, relative dispersion\cite{Larcheveque}, homogeneous shear\cite{Cambon81,Berto81}, isotropic turbulence with a mean scalar gradient\cite{Herr,Bos2005}, decay of turbulence in a wall bounded domain \cite{Touil2}, compressible turbulence\cite{Bataille}, 2D turbulence\cite{Pouquet}, premixed flame propagation.\cite{Ulitsky3} Without being exhaustive this list illustrates the variety of problems that have been adressed using this single-time closure. EDQNM was also used to propose subgrid models for Large-Eddy Simulation \cite{Chollet,Berto84}. One common feature of all these works is that they rely on the original heuristic assumption of EDQNM: the presence of an eddy damping term. 
The eddy damping is an essential ingredient of the EDQNM closure, as it represents the nonlinear damping of the triple correlations, necessary to obtain inertial range spectra in agreement with classical Kolmogorov \cite{Kolmogorov} theory. It should be pointed out that in the case of weakly nonlinear wave-turbulence only the presence of the damping is mandatory.\cite{CambonJacquin} A convenient specification of its form is of primary importance as soon as strong turbulence is considered. In the case of EDQNM, the Eddy-Damping is specified by dimensional analysis and an ad-hoc constant is introduced in the model. For other closures like the Test Field Model proposed by Kraichnan, \cite{KraichnanTFM} in which an auxiliary velocity field that does not respect the incompressibility constraint is introduced, the damping is deduced from a more sophisticated analysis, but an ad-hoc constant still has to be introduced (of order unity in the case of the Test Field Model).   

More elaborate are the two-point two-time closures proposed by Kraichnan (DIA, for Direct Interaction Approximation \cite{KraichnanDIA}), obtained by perturbation techniques applied to the Navier Stokes equations. 
The Eulerian formulation of DIA, violating the principle of statistical Galilean invariance and therefore being incompatible with a $K^{-5/3}$ Kolmogorov inertial range, was reformulated in a Lagrangian framework. The Lagrangian History DIA (LHDIA)\cite{Kraichnan65}, its abridged versions,\cite{Kraichnan65,KraichnanHerring} as well as the version of Kaneda \cite{Kaneda81} are known to yield predictions in agreement with Kolmogorov spectra without introducing any ad-hoc constant. The price one has to pay is the complexity of the models that depend on the entire Lagrangian history of the flow in contrast to the single-time theories. 

It is known that the EDQNM equations for isotropic turbulence can be obtained from two-time theories by assuming an exponential decay of both the response function and the two-time correlations. The eddy damping then corresponds to the inverse of the correlation time of the turbulent velocity field. In the framework of a Lagrangian formulation of the theories, this correlation time has to be defined along fluid particle trajectories and indeed a definition, corresponding to LHDIA for isotropic turbulence, can be found in Kraichnan \cite{Kraichnan65}:
\begin{eqnarray}\label{eq9.5}
\tau(K,t)=\int_{0}^t{\frac{E(K,t|s)}{E(K,t)}}ds
\end{eqnarray}
with $E(K,t)$ the energy spectrum and $E(K,t|s)$ the Lagrangian two-time energy spectrum (definitions are given below). 

In the EDQNM model, this time is modeled as a function of $E(K,t)$ and $K$, yielding by dimensional analysis:
\begin{equation}
(\tau(K,t))^{-1}=\eta(K,t)=\alpha \sqrt{ K^3 E(K,t)}
\end{equation} 
Another variant, non-local in wave-number space, is the expression proposed in Pouquet {\it et al.} \cite{Pouquet}:
\begin{equation}\label{eqpouq}
(\tau(K,t))^{-1}=\eta(K,t)=\lambda \sqrt{\int_0^K S^2 E(S,t)dS}.
\end{equation} 
At high Reynolds number, both formulations lead to
\begin{equation}
\tau(K,t)\sim K^{-2/3}
\end{equation}
in the inertial range of the spectra, mimicking the scaling that the Lagrangian correlation timescale is expected to follow in agreement with Kolmogorov theory. The constant $\alpha$ (or $\lambda$) is specified to obtain the desired value of the Kolmogorov constant.

The aim of the present letter is to derive a single-time  closure that does not contain ad-hoc specification of the damping term nor adjustable constant. The eddy damping timescale  will be determined within the closure, the key element of the model being the use of equation (\ref{eq9.5}). 

The Lagrangian two-time spectral tensor is defined by
\begin{equation}
\Phi_{ij}(\bm K,t|s)=FT_{{\bm x}-{\bm x}'}\left[\left<u_i({\bm x},t)u_j^L({\bm x}',t|s)\right>\right]\qquad (s\leq t),
\end{equation}
in which $FT_{{\bm x}-{\bm x}'}$ denotes a Fourier transform with respect to ${\bm x}-{\bm x}'$. $u_j^L({\bm x}',t|s)$ is defined as the velocity measured at time $s$ within the fluid element which passes through the point $\bm {\bm x}'$ at time $t$. In isotropic turbulence the two-time energy spectrum $E(K,t|s)$ is related to this tensor by the relation:
\begin{equation}
P_{ij}(\bm K)\frac{E(K,t|s)}{4\pi K^2}=\Phi_{ij}(\bm K,t|s)
\end{equation}
with $P_{ij}(\bm K)=(\delta_{ij}-K_iK_j/K^2)$. $E(K,t|s)$ is a key quantity as it appears in equation (\ref{eq9.5}) giving the timescale that has to be specified in the closure. It is a two-time Lagrangian quantity and is therefore difficult to evaluate in the framework of a one-time closure. However, only the time integral of $E(K,t|s)$ is required to express $\tau(K,t)$. This integral satisfies: 
\begin{eqnarray}\label{eqintuu}
\frac{P_{ij}(\bm K)}{4\pi K^2}\int_{0}^t{E(K,t|s)}ds=\nonumber\\
FT_{{\bm x}-{\bm x}'}\left[\left<u_i({\bm x},t)\int_{0}^t{u_j^L({\bm x}',t|s)}ds\right>\right].
\end{eqnarray}
The integral of the Lagrangian velocity along the trajectory in (\ref{eqintuu}) is the displacement of a fluid particle. Calling $a_j$ this displacement:
\begin{eqnarray}
a_j({\bm x},t)&=&X_j({\bm x},t|t)-X_j({\bm x},t|t=0)\nonumber\\
&=&\int_{0}^tu_j^L({\bm x},t|s)ds
\end{eqnarray}
and
\begin{eqnarray}
X_j({\bm x},t|t)=x_j,
\end{eqnarray}
(\ref{eqintuu}) can be written as:
\begin{eqnarray}
\frac{P_{ij}(\bm K)}{4\pi K^2}\int_{0}^t{E(K,t|s)}ds=\mathcal{F}_{u_ia_j}(\bm K,t),
\end{eqnarray}
with
\begin{eqnarray}\label{eqfKuiaj}
\mathcal{F}_{u_ia_j}(\bm K,t)=FT_{{\bm x}-{\bm x}'}[<u_i({\bm x},t)a_j({\bm x}',t)>].
\end{eqnarray}
In isotropic incompressible turbulence this quantity can be expressed as (Lumley \cite{Lumley2}):
\begin{eqnarray}\label{pijfua}
\mathcal{F}_{u_ia_j}(\bm K,t)=\frac{P_{ij}(\bm K)}{4\pi K^2}F_{ua}(K,t),
\end{eqnarray}
and therefore expression (\ref{eq9.5}) can be recasted as a function of one-time Eulerian quantities only:
\begin{eqnarray}\label{mjiklo}
\tau(K,t)=\frac{F_{ua}(K,t)}{E(K,t)} 
\end{eqnarray}
This expression forms the basis of the single-time two-point closure proposed in this paper. In the following, a way to obtain an expression for $F_{ua}(K,t)$ will be proposed that together with the equation for $E(K,t)$ and relation (\ref{mjiklo}) leads to a closed set of equations.

From the Navier-Stokes equations and the equation for the displacement of a fluid particle:
\begin{eqnarray}
\frac{d a_j({\bm x},t)}{dt}=u_j({\bm x},t),
\end{eqnarray}
a one-time two-point closure for $E(K,t)$ and $F_{ua}(K,t)$ in isotropic turbulence (and alternatively for $\Phi_{ij}(\bm K,t)$ and $\mathcal{F}_{u_ia_j}(\bm K,t)$ for anisotropic turbulence) can straightforwardly be written, by applying the Quasi-Normal approximation and Markovian assumption and expressing the relaxation time of the triple correlations using (\ref{mjiklo}). In the present letter, instead of deriving the evolution equation for $F_{ua}(K,t)$, we adopt a simpler approach, taking advantage of the analogy existing between the fluid particle displacement and an advected non-diffusive scalar field. This analogy will permit to express the closure model using only existing published equations. As pointed out by Batchelor,\cite{BatchelorEuler} a non-diffusive passive scalar in isotropic turbulence with a mean scalar gradient obeys the same equation as the displacement of a fluid particle. Considering an isotropic turbulence initially free from passive scalar fluctuations on which, at $t=0$, a mean scalar gradient is imposed in an arbitrary direction, $\partial \overline{\Theta}/\partial x_j$, the interaction of the velocity field with the scalar gradient  produces a scalar fluctuation $\theta$ governed by:
\begin{eqnarray}\label{eqbatch}
\frac{d \theta({\bm x},t)}{d t}=-\frac{\partial \overline{\Theta}}{\partial x_j} u_j({\bm x},t).
\end{eqnarray}
Integrating (\ref{eqbatch}) over the Lagrangian trajectory of the fluid particle that  arrives at time $t$ at position ${\bm x}$ yields: 

\begin{eqnarray}
\theta({\bm x},t)=-\frac{\partial \overline{\Theta}}{\partial x_j}(X_j({\bm x},t|t)-X_j({\bm x},t|t=0))\nonumber\\
= -\frac{\partial \overline{\Theta}}{\partial x_j}a_j({\bm x},t)
\end{eqnarray}
and the correlation between $\theta({\bm x}',t)$ and $u_i({\bm x},t)$ can be expressed as:
\begin{eqnarray}\label{eqgradxj}
<{u_i({\bm x},t)\theta({\bm x}',t)}>=-\frac{\partial \overline{\Theta}}{\partial x_j}<{u_i({\bm x},t)a_j({\bm x}',t)}>.
\end{eqnarray}
For isotropic turbulence without loss of generality, the direction of the gradient can arbitrarily be specified: for example $x_3$. Equation (\ref{eqgradxj}) then leads to:
\begin{eqnarray}\label{link}
F_{u_3\theta}(K,t)=-\frac{\partial \overline{\Theta}}{\partial x_3}F_{u_3a_3}(K,t)
\end{eqnarray}
with $F_{u_3\theta}(K,t)$ defined as the scalar flux spectrum (see for example O'Gorman and Pullin \cite{Gorman}), or introducing $F_{ua}(K,t)$ as in equation (\ref{mjiklo}):
\begin{eqnarray}\label{link2}
F_{u_3\theta}(K,t)=-\frac{2}{3}\frac{\partial \overline{\Theta}}{\partial x_3}F_{ua}(K,t).
\end{eqnarray}
since isotropy implies that $F_{u_1a_1}=F_{u_2a_2}=F_{u_3a_3}=\frac{2}{3}F_{ua}$.

One can also arbitrarily specify the value of the gradient, since the scalar equation is linear. Choosing $\partial \overline{\Theta}/\partial x_3=-3/2$ simplifies the formulation. With this particular value of the gradient, equation (\ref{link2}) simply expresses the identity between the spectrum of the scalar flux of the non-diffusive scalar and the spectrum of the velocity-displacement correlation. Hence it is straightforward to use the EDQNM model proposed by Herr \etal \cite{Herr} or Bos \etal~ \cite{Bos2005} for the scalar flux spectrum to calculate $F_{u_3\theta}(K,t)=F_{ua}(K,t)$ (in this work we use the formulation of Bos \etal). The equations for $E(K,t)$ and $F_{ua}(K,t)$ will be solved simultaneously to calculate the energy spectrum in isotropic turbulence and to evaluate the damping term using (\ref{link}) that now takes the form:
\begin{equation}\label{trotro}
\eta(K,t)=(\tau(K,t))^{-1}=\frac{E(K,t)}{F_{u_3\theta}(K,t)}.
\end{equation}

The evolution equation for the energy spectrum is:
\begin{eqnarray}\label{eqLin}
\left[\frac{\partial }{\partial t}+2\nu K^2\right]E(K,t)=T_{NL}(K,t),
\end{eqnarray}
in which the expression for the nonlinear transfer $T_{NL}$ is the classical single-time two-point closure expression (Orszag \cite{Orszag}):
\begin{eqnarray}\label{eqTnlclassic}
T_{NL}(K,t)=\iint_{\Delta}\Theta(K,P,Q)~(xy+z^3)\left[K^2PE(P,t)E(Q,t) \right.\nonumber\\
\left. -P^3E(Q,t)E(K,t)\right]\frac{dPdQ}{PQ},
\end{eqnarray}
where $\Delta$ is a band in $P,Q$-space so that the three wave vectors ${\bm{K}, \bm{P}, \bm{Q}}$ form a triangle. $x,y,z$ are the cosines of the angles opposite to the sides $K,P,Q$ of the triangle formed from  ${\bm{K}, \bm{P}, \bm{Q}}$. The characteristic time $\Theta(K,P,Q)$ is defined as:
\begin{equation}
\Theta(K,P,Q)=\frac{1-exp(-\mu_{KPQ}\times t)}{\mu_{KPQ}}
\end{equation}
with
\begin{equation}
\mu_{KPQ}=\nu(K^2+P^2+Q^2)+\eta(K,t)+\eta(P,t)+\eta(Q,t)
\end{equation}
The difference with the EDQNM model\cite{Orszag} is that the eddy damping in the equations is not anymore heuristically specified, but is calculated using relation (\ref{trotro}). 

The equation for $F_{u_3\theta}(K,t)$ is:
\begin{eqnarray}\label{eqFwtheta}
\left[\frac{\partial}{\partial t}+\nu K^2\right]F_{u_3\theta}(K,t)=
P(K,t)+T^{NL}_{u_3\theta}(K,t)+\Pi(K,t),
\end{eqnarray}
which is the equation of reference (12) in the particular case of a non-diffusive scalar. In (\ref{eqFwtheta}), $P(K,t)$ is a term that in this case can be interpreted as the production of scalar flux by the mean scalar gradient, such that $P(K,t)=-\frac{2}{3}\frac{\partial \overline{\Theta}}{\partial x_3} E(K,t)=E(K,t)$. The expressions for the nonlinear terms $T^{NL}_{u_3\theta}(K,t)$ and $\Pi(K,t)$ are not reproduced here (Equations (14) and (15) of reference (12)). These closed terms are exactly the same as in the case of the EDQNM model, except that the eddy damping is determined by equation (\ref{trotro}). More specifically, in relation (16) of reference (12),
\begin{equation}\label{mufkpq}
\mu^F_{KPQ}=\mu'(K)+\mu'(P)+\mu''(Q)+\nu(K^2+P^2),
\end{equation}
$\mu'$ is replaced by $\eta$ and $\mu''$ is still zero as in Bos \cite{Bos2005}.

\begin{figure} 
\begin{center}
\includegraphics[width=.39\textwidth]{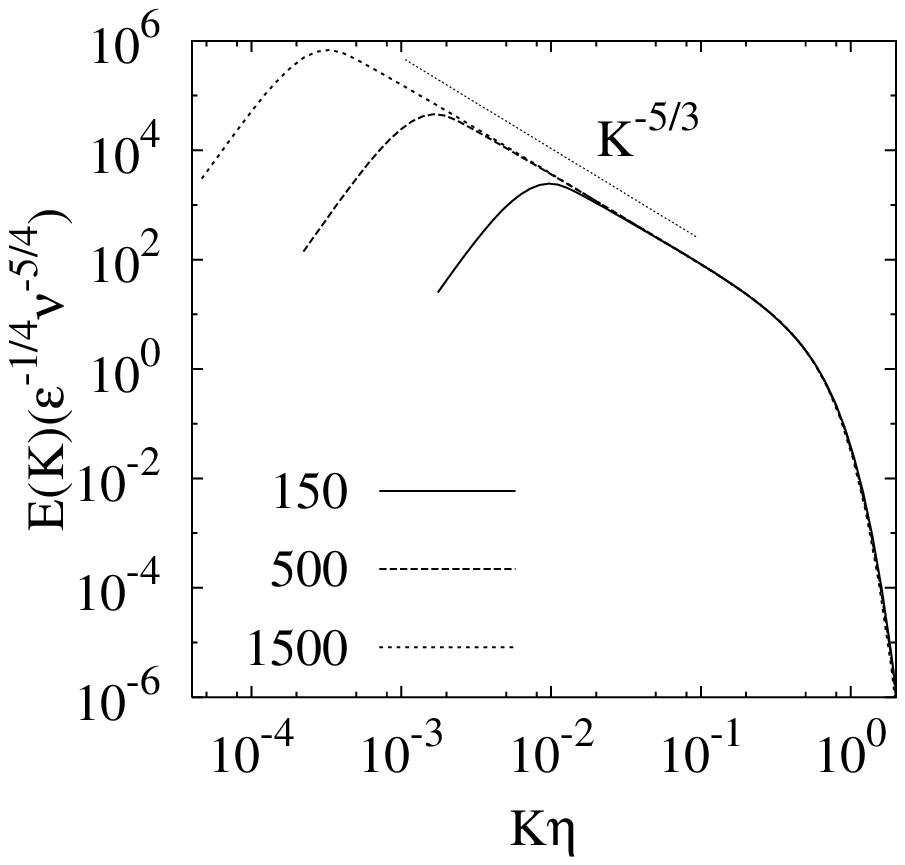}\\
\includegraphics[width=.39\textwidth]{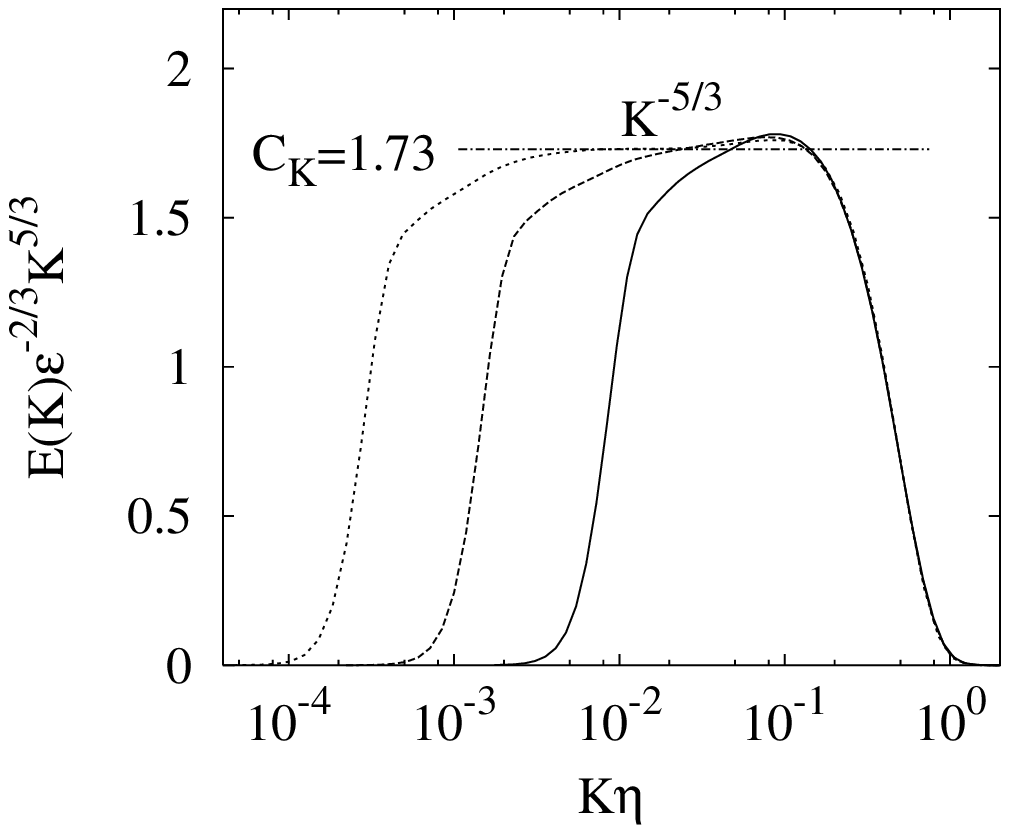}
 \end{center}
\caption{\label{EKBertodamped}  Top: energy spectrum  at $R_\lambda=150$, $500$ and $1500$; bottom: compensated form.}
\end{figure}

The model is applied to the decay of isotropic turbulence by numerically integrating equations (\ref{eqLin}) and (\ref{eqFwtheta}). The energy spectrum is initialized by:
\begin{equation}\label{initexp}
E(K,0)=B K^4e^{-2K^2/K_L^2},
\end{equation}
with $K_0=1$, $K_L=10$ and  $B$ determined so that the initial kinetic energy is equal to $1$. The energy spectrum was evaluated during the period of self-similar decay. Spectra are shown at $R_\lambda=150$, $500$ and $1500$. The results in figure \ref{EKBertodamped} show that a $K^{-5/3}$ inertial range is obtained for the energy spectrum. The value of the Kolmogorov constant is estimated to be $1.73$ as can be seen when the spectrum is shown in compensated form. It has to be reminded that in the case of the EDQNM closure this value is not a prediction of the model but has to be specified by choosing the constant $\lambda$. The value of $C_K=1.73$ can be obtained with EDQNM by choosing the value $\lambda=0.49$ in expression (\ref{eqpouq}). A detailed comparison between the EDQNM model and the present closure deserves further attention.

The results in figure \ref{EKBertodamped} suggest that the present model yields a reasonable estimate of the Lagrangian timescale in isotropic turbulence. It would be useful to compare the present results to estimations of the timescale provided by Direct Numerical Simulation using the method proposed in Lee \etal \cite{Lee} (see also Gotoh \etal \cite{Gotoh2}) or to higher Reynolds number data provided by Large Eddy Simulation.\cite{Rubinstein2}

\begin{thebibliography}{10}

\bibitem{Orszag}
S.A. Orszag.
\newblock Analytical theories of turbulence.
\newblock { J. Fluid Mech.}, {\bf 41}, 363, (1970).

\bibitem{Leith2}
C.E. Leith.
\newblock Atmospheric predictability and two-dimensional turbulence.
\newblock { J. Atmos. Sci.}, {\bf 28}, 145, (1971).

\bibitem{VignonPOF}
J.-M. Vignon and C.~Cambon.
\newblock Thermal spectral calculation using eddy-damped quasi-normal markovian
  theory.
\newblock { Phys. Fluids}, {\bf 23}, 1935, (1980).

\bibitem{Herring}
J.R. Herring, D.~Schertzer, M.~Lesieur, G.R. Newman, J.P. Chollet, and
  M.~Larcheveque.
\newblock A comparative assessment of spectral closures as applied to passive
  scalar diffusion.
\newblock { J. Fluid Mech.}, {\bf 124}, 411, (1982).

\bibitem{CambonJacquin}
C.~Cambon and L.~Jacquin.
\newblock Spectral approach to non-isotropic turbulence subjected to rotation.
\newblock { J. Fluid Mech.}, {\bf 202}, 295, (1989).

\bibitem{GodeferdPOF}
F.S. Godeferd and C.~Cambon.
\newblock Detailed investigation of energy transfers in homogeneous stratified
  turbulence.
\newblock { Phys. Fluids}, {\bf 6}, 2084, (1994).

\bibitem{PouquetMHD}
U.~Frisch, A.~Pouquet, J.~Leorat, and A.~Mazure.
\newblock Possibility of an inverse cascade of magnetic helicity in
  magnetohydrodynamic turbulence.
\newblock { J. Fluid Mech.}, {\bf 68}, 769, (1975).

\bibitem{Larcheveque}
M.~Larcheveque and M.~Lesieur.
\newblock The application of eddy-damped markovian closures to the problem of
  dispersion of particle pairs.
\newblock { J. M{\'e}canique}, {\bf 20}, 113, (1981).

\bibitem{Cambon81}
C.~Cambon, D.~Jeandel, and J.~Mathieu.
\newblock Spectral modelling of homogeneous non-isotropic turbulence.
\newblock { J. Fluid Mech.}, {\bf 104}, 247, (1981).

\bibitem{Berto81}
J.P. Bertoglio.
\newblock A model of three-dimensional transfer in non-isotropic homogeneous
  turbulence.
\newblock In { Turbulent Shear Flows {III}, Springer-Verlag}, 253,
   (1981).

\bibitem{Herr}
S.~Herr, L.P. Wang, and L.R. Collins.
\newblock {EDQNM} model of a passive scalar with a uniform mean gradient.
\newblock { Phys. Fluids}, {\bf 8}, 1588, (1996).

\bibitem{Bos2005}
W.J.T. Bos, H.~Touil, and J.-P. Bertoglio.
\newblock Reynolds number dependency of the scalar flux spectrum in isotropic
  turbulence with a uniform scalar gradient.
\newblock { Phys. Fluids}, {\bf 17}, 125108  (2005).

\bibitem{Touil2}
H.~Touil, L.~Shao, and J.P. Bertoglio.
\newblock The decay of turbulence in a bounded domain.
\newblock { J. Turbul.}, {\bf 3}, 049, (2002).

\bibitem{Bataille}
J.P. Bertoglio, F.~Bataille, and J.D. Marion.
\newblock Two-point closures for weakly compressible turbulence.
\newblock { Phys. Fluids}, {\bf 13}, 290, (2001).

\bibitem{Pouquet}
A.~Pouquet, M.~Lesieur, J.C. Andr{\'e}, and C.~Basdevant.
\newblock Evolution of high {R}eynolds number two-dimensional turbulence.
\newblock { J. Fluid Mech.}, {\bf 72}, 305, (1975).

\bibitem{Ulitsky3}
M.~Ulitsky and L.R. Collins.
\newblock Application of the eddy-damped quasi-normal markovian spectral
  transport theory to premixed turbulent flame propagation.
\newblock { Phys. Fluids}, {\bf 9}, 3410, (1996).

\bibitem{Chollet}
J.P. Chollet and M.~Lesieur.
\newblock Parameterization for small scales of three-dimensional isotropic
  turbulence using spectral closures.
\newblock { J. Atmos. Sci.}, {\bf 38}, 2747, (1981).

\bibitem{Berto84}
J.-P. Bertoglio and J.~Mathieu.
\newblock Mod\'elisation stochastique des petites \'echelles de la turbulence:
  formulation g\'en\'eral.
\newblock { Compte-Rendu Ac. Sci},  {II}-12, 751, (1984).

\bibitem{Kolmogorov}
A.N. Kolmogorov.
\newblock The local structure of turbulence in incompressible viscous fluid for
  very large {R}eynolds numbers.
\newblock { Dokl. Akad. Nauk. SSSR}, {\bf 30}, 301, (1941).

\bibitem{KraichnanTFM}
R.H. Kraichnan.
\newblock An almost-markovian galilean-invariant turbulence model.
\newblock {J. Fluid Mech.}, {\bf 47}, 513, (1971).

\bibitem{KraichnanDIA}
R.H. Kraichnan.
\newblock The structure of isotropic turbulence at very high {R}eynolds
  numbers.
\newblock { J. Fluid Mech.}, {\bf 5}, 497, (1959).

\bibitem{Kraichnan65}
R.H. Kraichnan.
\newblock Lagrangian-history closure approximation for turbulence.
\newblock { Phys. Fluids}, {\bf 8}, 575, (1965).

\bibitem{KraichnanHerring}
R.H. Kraichnan and J.R. Herring.
\newblock A strain-based lagrangian-history turbulence theory.
\newblock { J. Fluid Mech.}, {\bf 88}, 355, (1978).

\bibitem{Kaneda81}
Y.~Kaneda.
\newblock Renormalized expansions in the theory of turbulence with the use of
  the lagrangian position function.
\newblock { J. Fluid. Mech.}, {\bf 107}, 131, (1981).

\bibitem{Lumley2}
J.L. Lumley.
\newblock The spectrum of nearly inertial turbulence in a stably stratified
  fluid.
\newblock { J. Atmos. Sci.}, {\bf 21}, 99, (1964).

\bibitem{BatchelorEuler}
G.K. Batchelor.
\newblock Diffusion in a field of homogeneous turbulence. {I}. eulerian
  analysis.
\newblock { Aust. J. Sci. Res. Ser. A}, {\bf 2}, 437, (1949).

\bibitem{Gorman}
P.A. O'Gorman and D.I. Pullin.
\newblock The velocity-scalar cross correlation of stretched spiral vortices.
\newblock {Phys. Fluids}, {\bf 15}, 280, (2003).

\bibitem{Lee}
C.H. Lee, K.~Squires, J.-P. Bertoglio, and J.~Ferziger.
\newblock Study of lagrangian characteristic times using direct numerical
  simulation of turbulence.
\newblock In { Turbulent Shear Flows {VI}, Toulouse}, Springer-Verlag, 58,
   (1987).

\bibitem{Gotoh2}
T.~Gotoh, R.S. Rogallo, J.R. Herring, and R.H. Kraichnan.
\newblock Lagrangian velocity correlations in homogeneous isotropic turbulence.
\newblock { Phys. Fluids}, {\bf 5}, 2846, (1993).

\bibitem{Rubinstein2}
Guo-Wei He, R.~Rubinstein, and Lian-Ping Wang.
\newblock Effects of subgrid-scale modeling on time correlations in large eddy
  simulation.
\newblock { Phys. Fluids}, {\bf 14}, 2186, (2002).

\end{thebibliography}
\end{document}